%% file: main.tex
\PassOptionsToPackage{dvipsnames}{xcolor}
\documentclass[man,12pt,floatsintext,donotrepeattitle,biblatex]{apa7}

\newif\ifcorrections
\correctionstrue 
\newcommand{\fix}[2]{\ifcorrections#2\else#1\fi}
\newif\ifcomments
\commentsfalse

\usepackage[american]{babel}
\usepackage[T1]{fontenc}

\usepackage[scaled=1.0135]{newtxtext}      
\usepackage{newtxmath}
\usepackage{microtype}
\usepackage{needspace}
\microtypesetup{protrusion=false,expansion=false}
\usepackage{csquotes}

\DeclareLanguageMapping{american}{american-apa}
\AtEveryCitekey{\iffieldequalstr{entrykey}{yun2026hybrid}{\defcounter{minnames}{2}}{}}

\usepackage{graphicx}
\graphicspath{{figures/}}
\usepackage{booktabs}
\usepackage{array}
\usepackage{xcolor}
\usepackage{tikz}
\usetikzlibrary{positioning, shapes.multipart, arrows.meta}
\usepackage{listings}

\definecolor{codebg}{HTML}{FAFAFA}
\definecolor{codetag}{HTML}{E45649}
\definecolor{codeattr}{HTML}{C18401}
\definecolor{codestr}{HTML}{50A14F}
\definecolor{codekw}{HTML}{A626A4}
\definecolor{codefn}{HTML}{4078F2}
\definecolor{codevar}{HTML}{E06C75}
\definecolor{codeop}{HTML}{0184BC}
\definecolor{codetext}{HTML}{383A42}
\definecolor{linkblue}{HTML}{1155CC}

\newcommand{\codefont}{\fontfamily{DejaVuSansMono-TLF}\selectfont}
\lstdefinestyle{html}{
  language=HTML,
  tagstyle=\color{codetag},
  keywordstyle=\color{codeattr},
  morekeywords={width,height,src},
  stringstyle=\color{codestr},
  usekeywordsintag=true,
  markfirstintag=false,
}
\newcommand{\CodeFile}[2][]{\par\noindent\begin{minipage}{\linewidth}%
  \lstinputlisting[aboveskip=6pt,belowskip=6pt,#1]{#2}\end{minipage}\par
  \global\everypar{{\setbox0\lastbox}\global\everypar{}}}
\lstdefinestyle{python}{
  language=Python,
  keywordstyle=\color{codekw},
  keywords={import,try,except,as},
  keywordstyle=[2]\color{codetag},
  keywords=[2]{print},
  emph={get,raise_for_status,json,replace,exceptions,RequestException},
  emphstyle=\color{codefn},
  identifierstyle=\color{codevar},
  stringstyle=\color{codestr},
  morestring=[b]""",
  literate={=}{{\textcolor{codeop}{=}}}1,
}

\usepackage{xurl}                         
\usepackage{hyperref}
\hypersetup{colorlinks=true, linkcolor=black, citecolor=black, urlcolor=linkblue}
\AtBeginBibliography{\hypersetup{urlcolor=black}}
\AtBeginBibliography{\mathchardef\UrlBreakPenalty=9000 \mathchardef\UrlBigBreakPenalty=9000 }
\DeclareFieldFormat{url}{\url{#1}\nopunct}
\DeclareFieldFormat[article]{eid}{\bibcpstring{jourarticle}\space#1}

\ExplSyntaxOn
\NewDocumentCommand{\codestr}{m}{
  \tl_set:Ne \l_tmpa_tl { \detokenize{#1} }
  \tl_replace_all:Nnn \l_tmpa_tl { - } { -\kern0pt }
  \tl_use:N \l_tmpa_tl
}
\ExplSyntaxOff
\newcommand{\code}[1]{{\fontsize{10}{0}\selectfont\texttt{\codestr{#1}}}}   
\newcommand{\Note}[1]{\par\noindent\textit{Note}. #1\par} 
\newcommand{\FigureTitle}[3]{\par\noindent\refstepcounter{#1}\label{#2}%
  \textbf{\ifnum\pdfstrcmp{#1}{figure}=0 Figure\else Table\fi~\csname the#1\endcsname}\par\nopagebreak
  \noindent\textit{#3}\par}
\newdimen\titledepth
\newdimen\blockheight
\newcommand{\FigureBlock}[3]{\par\titledepth=\prevdepth\nopagebreak\nointerlineskip
  \vskip\dimexpr 8.28pt-\titledepth\relax
  \setbox0\vbox{\vskip #2\relax #3}\blockheight=#1\relax
  \ifdim\dimexpr\ht0+\dp0\relax>\blockheight \blockheight=\dimexpr\ht0+\dp0\relax\fi
  \vbox to \blockheight{\unvbox0\vss}\prevdepth=0pt\relax}
\newcommand{\NewPageUnlessBoxFits}[1]{\par\penalty0
  \ifdim\dimexpr\pagegoal-\pagetotal\relax<\dimexpr\ht#1+\dp#1+80pt\relax\newpage\fi}
\newcommand{\BoxBlock}[1]{\FigureBlock{\dimexpr\ht#1+\dp#1+6.9pt\relax}{6.9pt}{\copy#1}}
\newcommand{\CodeBox}[3]{{\setlength{\fboxsep}{0pt}\colorbox{codebg}{\parbox[c][#2][c]{#1}{\hspace*{9.5pt}\begin{minipage}{\dimexpr#1-19pt\relax}\lstset{frame=none,backgroundcolor={},xleftmargin=0pt,xrightmargin=0pt}#3\end{minipage}}}}}
\newcommand{\codeV}[1]{\code{#1}\vadjust{\vskip2.2pt}}
\newcommand{\TableSetup}{\tabletext\setlength{\tabcolsep}{5.2pt}\renewcommand{\arraystretch}{0.4}%
  \setlength{\heavyrulewidth}{1.33pt}\setlength{\lightrulewidth}{1.33pt}%
  \setlength{\aboverulesep}{0pt}\setlength{\belowrulesep}{0pt}\setlength{\abovetopsep}{0pt}}
\newcommand{\HeadRow}{\rule[-1pt]{0pt}{10.7pt}}      
\newcommand{\Row}[1][11.5pt]{\rule[-1.53pt]{0pt}{#1}} 
\newcommand{\blankline}{\par\noindent\mbox{}\par} 
\usepackage{framed}
\newcounter{box}
\newenvironment{InfoBox}[2]{\par\medskip\begin{framed}\singlespace\refstepcounter{box}\label{#1}%
  \noindent\textbf{Box~\thebox}\par\noindent\textit{#2}\par\smallskip}{\end{framed}\medskip}

\newcommand{\bodyspread}{1.9114}   
\AtBeginDocument{\linespread{\bodyspread}\selectfont\frenchspacing\setlength{\parindent}{0.5in}}
\usepackage{indentfirst}                 
\makeatletter
\fancypagestyle{titlepage}{\fancyhf{}\fancyhead[L]{\MakeUppercase{\@shorttitle}}\fancyhead[R]{\thepage\hspace*{6pt}}}
\fancypagestyle{otherpage}{\fancyhf{}\fancyhead[L]{\MakeUppercase{\@shorttitle}}\fancyhead[R]{\thepage\hspace*{6pt}}}
\renewcommand\section[1]{\par\Needspace{3\baselineskip}\phantomsection\addcontentsline{toc}{section}{#1}{\centering\normalfont\bfseries #1\par}}
\renewcommand\subsection[1]{\par\Needspace{3\baselineskip}\phantomsection\addcontentsline{toc}{subsection}{#1}{\noindent\normalfont\bfseries #1\par}}
\renewcommand\subsubsection[1]{\par\Needspace{3\baselineskip}\phantomsection\addcontentsline{toc}{subsubsection}{#1}{\noindent\normalfont\bfseries\itshape #1\par}}
\makeatother

\newcommand{\tabletext}{\linespread{1}\fontsize{10}{11.5}\selectfont} 
\newcommand{\singlespace}{\linespread{1}\selectfont}
\makeatletter
\renewcommand\footnoterule{\kern-3\p@\hrule\@width 144pt\kern0.9pt}
\makeatother
\let\origfootnote\footnote
\renewcommand{\footnote}[1]{\origfootnote{\linespread{1}\fontsize{12}{13.8}\selectfont #1}}

\title{Facilitating the Integration of LLMs Into Online Experiments With Simple Chat}
\shorttitle{Facilitating the Integration of LLM\lowercase{s}}

\begin{document}

\thispagestyle{titlepage}
\newcommand{\centredline}[1]{\noindent\makebox[\linewidth]{#1}\par}
\centredline{}
\centredline{}
\centredline{\textbf{Facilitating the Integration of LLMs Into Online Experiments With Simple Chat}}
\centredline{}
\centredline{Rodrigo Bermudez Schettino\textsuperscript{1}, Ali Dasmeh\textsuperscript{1}, and Levin Brinkmann\textsuperscript{1}}
\centredline{\textsuperscript{1}\,Center for Humans and Machines, Max Planck Institute for Human Development}
\centredline{}\centredline{}\centredline{}\centredline{}\centredline{}\centredline{}\centredline{}\centredline{}
\centredline{\textbf{Author Note}}
No funding was received to assist with the preparation of this manuscript. We have no conflicts of interest to disclose. Correspondence concerning this article should be addressed to Rodrigo Bermudez Schettino, Center for Humans and Machines, Max Planck Institute for Human Development, Lentzeallee 94, Berlin, 14195, Germany. Telephone: +49 30 82406-0. Email: schettino@mpib-berlin.mpg.de
\clearpage

{\fontsize{11}{13.29}\selectfont
{\centering\textbf{Abstract}\par}
\noindent As large language models (LLMs) become increasingly prevalent, understanding human--LLM interactions is emerging as a central priority in psychological research. Online experiments offer an efficient means to study human--LLM interactions, yet integrating LLMs into established survey platforms remains technically demanding, particularly when aiming for ecologically realistic, real-time conversational experiences with strong experimental control. We introduce Simple Chat, an open-source, research-focused chat interface that streamlines LLM integration for platforms such as Qualtrics, oTree, and LimeSurvey, while presenting a unified participant experience across conditions. Simple Chat connects to both commercial providers and open-weights models, displays streamed responses incrementally, and offers an administrative interface for fine-grained control of prompts and interface features. Simple Chat is available both as a hosted service that requires no installation and as open-source software that researchers can deploy on their own server or adapt to their needs. By reducing technical barriers and standardizing interfaces, Simple Chat helps advance the study of human--LLM interaction. In this article, we outline Simple Chat's key features, provide a step-by-step tutorial, and demonstrate its utility through two illustrative case studies.

\noindent\hspace*{30.9pt}\textit{Keywords:} online experiments, human-machine interaction, large language models, chatbots\par}
\clearpage

\section{Introduction}

Large language models (LLMs) have rapidly emerged as one of the most transformative advancements in artificial intelligence, bridging the gap between human communication and machine interpretation. Trained on vast amounts of data, these systems can interpret context, reason through complex ideas, and generate human-like language with remarkable fluency \parencite{chang2024language,jones2025people}. Their capabilities extend far beyond simple text generation---they are applied across a wide range of tasks, from brainstorming
\parencite[e.g.,][]{chang2025framework,fukumura2025can}
to programming \parencite[e.g.,][]{agarwal2024copilot,nam2024using}.

As a tool, large language models are increasingly integrated into psychological research, where they assist in generating both static and adaptive study materials, serve as stand-ins for human participants by simulating diverse responses and reproducing established findings, and streamline data-analytic workflows through automated coding, data cleaning, and statistical analysis, thereby expanding methodological possibilities and reducing labor demands \parencite{bail2024can,behrend2025participant,charness2025next,demszky2023using}.

Importantly, LLMs have also emerged as subjects of inquiry, particularly regarding how humans interact with them. Early studies examine human--LLM dynamics in domains such as community-based content moderation \parencite{mohammadi2025ai}, countering misinformation \parencite{rani2025can}, reducing belief in conspiracy theories \parencite{costello2024durably}, improving online political discourse \parencite{argyle2023leveraging}, and alleviating political polarization \parencite{walter2025using}. Experimental designs for studying human--LLM interactions can be placed on a spectrum of sophistication. At the most basic level, participants are directed to an external public platform to interact with the model before returning to the study. While easy to implement, this approach offers limited control over the interaction process. A more advanced approach involves embedding the LLM directly into the experimental environment, allowing structured dialogues that can be recorded, analyzed, and tailored to the research design. This approach provides richer insights into human-AI interaction while maintaining experimental control. At the most advanced level, LLMs can be configured to assume specific roles or personas, enabling researchers to systematically shape social interactions in ways previously achievable only with trained actors \parencite{behrend2025participant}. Additionally, LLMs can also be employed to conduct qualitative research and interview people at scale \parencite{skeggs2025micro}.

\subsection{Existing Methods and Challenges of Integrating LLMs Into Online Experiments}
\label{sec:challenges}

Online experiments offer a fast and cost-effective way to conduct such research, which makes them well-suited for studying human--LLM interaction at scale \parencite{anwyl2020gorilla,rodd2024moving}. However, embedding LLMs into online experiments still poses considerable challenges. Popular platforms like Qualtrics lack native support for LLM integration.\footnote{It should be noted that \textcite{otree2025version} has introduced support for ``web APIs'' that could be used to query LLMs in version 6.0. However, it could introduce race conditions. Moreover, researchers must build the chat interface and save conversation history themselves. While this update shows that popular platforms are starting to realize the need for LLM integration features, such features are still at an early stage and may not yet be stable or robust.} As a consequence, researchers are dependent on custom workarounds that often involve contacting REST APIs via HTTP requests. While API-based approaches for specific platforms have been successfully implemented
\fix{(e.g., \citeauthor{costello2024durably} \citeyear{costello2024durably})}{\parencite[e.g.,][]{costello2024durably}},
they remain technically demanding and are thus largely inaccessible for researchers without extensive programming expertise.

Moreover, existing solutions are often restricted to specific platforms or providers. For example, \textcite{costello2024durably} created a template that can be used exclusively within Qualtrics. Similarly, \textcite{mckenna2023otree}'s oTree GPT is limited to oTree. oTree GPT also only allows integration with GPT models, as does GPT for Researchers (G4R; \cite{kim2025capture}). SCALE \parencite{valet2025scale} moves the model calls of oTree studies to a serverless function on Amazon Web Services (AWS) to work around oTree's limited handling of concurrent requests, but it requires an AWS account and is limited to OpenAI's models. LEXI \parencite{laban2024lexi} is a standalone platform that hosts its own questionnaires rather than extending an existing survey platform. The section Comparison With Existing Tools compares these tools with Simple Chat in detail. Some tools are also tailored to particular use cases, such as the one developed by
\fix{Chopra et al.\ (\citeyear{chopra2023conducting})}{\textcite{chopra2023conducting}},
which is designed specifically for LLM applications in interview settings.

Another challenge lies in the lack of interface standardization. Custom-built solutions developed by different researchers often result in variations in chat interface design. These inconsistencies can undermine comparability across studies \parencite{laban2024lexi}.

API latency, the time it takes for the API to respond, can also be a problem. Response delays can disrupt conversational flow and worsen users' experience with LLM-powered agents \parencite{maslych2025mitigating}.

This article is a practical tutorial whose primary objective is to enable researchers to integrate LLMs into their own online experiments with Simple Chat. It explains how to use the tool and highlights methodological considerations that arise in doing so. The remainder of the article is organized as follows:

\begin{itemize}
  \item \textbf{Simple Chat} introduces the tool, its deployment options, performance, and relationship to existing tools.
  \item \textbf{Quick Start} guides readers from registration to exporting a test conversation.
  \item \textbf{Case Studies} show how two published online experiments \parencite{costello2024durably,walter2025using} can be re-implemented with Simple Chat.
  \item \textbf{Discussion} addresses ethical considerations and limitations.
\end{itemize}
\blankline

\section{Simple Chat}
\label{sec:simple-chat}

In response to these issues, we developed Simple Chat, an HTML-based interface that empowers researchers to seamlessly integrate interactions with LLMs into their online experiments. Simple Chat can be embedded as an iframe within existing experimental setups. It thus provides a standardized interface for interactions with LLMs that functions independently of the host platform. At present, Simple Chat accommodates text and image inputs with text outputs, and its architecture can be readily extended to additional modalities such as audio or image generation. Notably, Simple Chat connects with a range of LLM providers, including OpenAI, Azure OpenAI, Anthropic, and Google. Generally speaking, Simple Chat supports any LLM provider with an OpenAI-compatible API: researchers set the provider's endpoint and add the model's name to Simple Chat's list of models. This includes LLMs hosted on institutional clusters. Figure~\ref{fig:architecture} provides a schematic overview of the Simple Chat architecture, illustrating how it links experimental platforms with LLM providers.

\FigureTitle{figure}{fig:architecture}{Overview of Simple Chat's Bridging Architecture Between Survey Platforms and LLM Providers}
\FigureBlock{183.4pt}{10.8pt}{\hbox{\hspace*{4.4pt}\includegraphics[width=371pt,height=139.8pt]{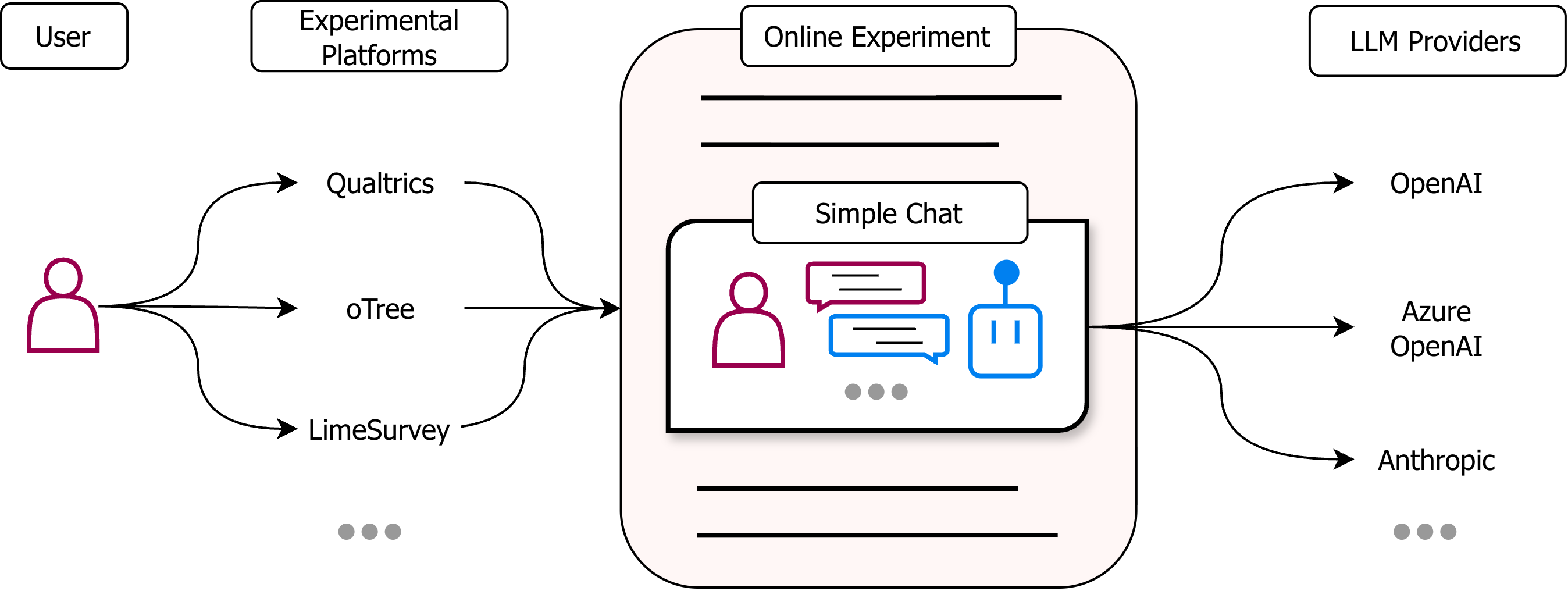}}}

Figure~\ref{fig:chat-interface} shows the Simple Chat interface. For the sake of external validity and to lower usage barriers for study participants, it is designed to resemble familiar LLM-based chatbot environments. The upper portion displays the complete dialogue history, including all previous human and LLM messages, and the lower portion features a text box for participants to compose and edit their replies. Once a study participant sends a message, it is added to the conversation thread. The submit button then remains disabled until the LLM generates its next response. Without streaming, the first characters of this response appeared on screen after a median of 1.3 to 3.1~s in our benchmark, depending on the model (see Performance below). Streaming displays responses incrementally, allowing participants to see text before generation is complete. Response onset and token pacing can shape users' experience of LLM interactions \parencite{tan2026impact,zhu2026just}.

\FigureTitle{figure}{fig:chat-interface}{Chat Interface}
\FigureBlock{265.5pt}{1pt}{\hbox{\hspace*{1.5pt}\setlength{\fboxsep}{0pt}\setlength{\fboxrule}{0.75pt}\fbox{\includegraphics[width=256pt,height=233.5pt]{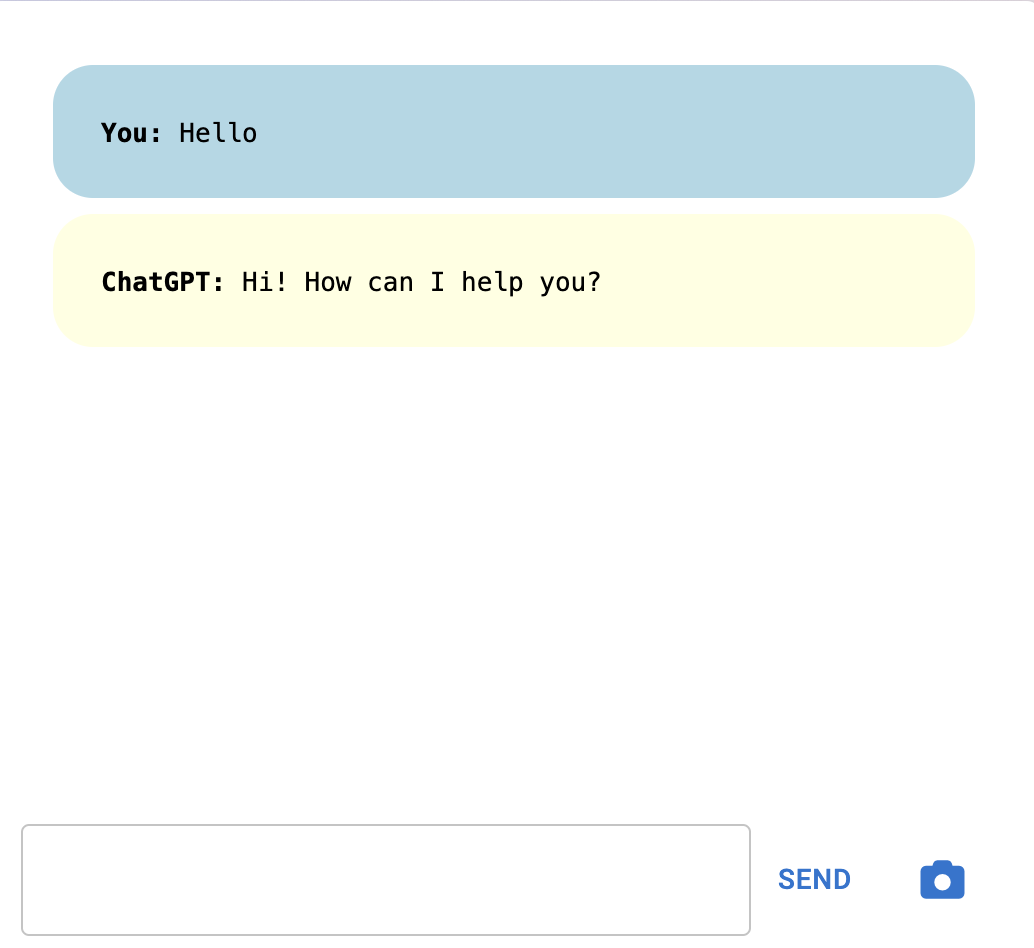}}}}

Simple Chat is well suited for online experiments across a wide range of research domains, including behavioral economics, cognitive psychology, and computational social science. It can also serve as a valuable tool for AI safety research. One particularly important application in this area is red-teaming, where model behavior is systematically evaluated under adversarial probing. With Simple Chat, the target LLM can be embedded directly into an online study setup. Researchers can thus follow up with standardized survey questions after participants probe the model, which allows them to collect detailed, structured data that can guide mitigation strategies and improve model safety.

\subsection{Hosted Service and Self-Hosting}
\label{sec:hosting-options}

Simple Chat is available as a hosted service and as open-source software. The hosted service, Simple Chat as a Service, suits any study that does not require changes to the code: researchers configure a project in the dashboard and embed the chat in their survey as an iframe, without setting up or maintaining a server. Researchers who want to run Simple Chat on their own infrastructure or adapt its code can deploy the open-source version on a server, which requires technical expertise (see Setup and Deployment). In both cases, Simple Chat replaces study-specific integration code with a reusable component.

Simple Chat as a Service is operated by the Center for Humans and Machines at the Max Planck Institute for Human Development (\url{https://www.mpib-berlin.mpg.de/2382048/Simple-Chat_-Reducing-Technical-Barriers-to-Human_LLM-Experiments}) and is open to any researcher. Researchers register with a verified email address, create a project in the dashboard, and enter an API key from their own LLM provider account, which is billed directly for model usage. The Quick Start section walks through these steps, and the service's documentation (\url{https://simple-chat.chm.mpib-berlin.mpg.de/docs/getting-started/quick-start/}) covers further settings.

\subsection{Addressing LLM Integration Challenges}
\label{sec:addressing}

Simple Chat comes with a range of features that help researchers integrate LLMs into online experiments more smoothly and effectively. It allows them to configure experimental parameters such as model type, system message, and session identifiers via query strings, which eliminates the need for custom-coded API integrations. Once the iframe is embedded, Simple Chat manages all model communication and session handling directly. This substantially lowers the technical barrier for researchers with limited programming expertise and reduces experimental setup time, because the same deployment can be reused across studies and survey platforms.

Simple Chat also reduces reliance on platform-specific implementations and eliminates constraints such as hard-coded conversation lengths. Unlike existing templates for platforms like Qualtrics, which are limited to a single platform, Simple Chat provides a modular solution that can be deployed across multiple platforms, including Qualtrics, oTree, and LimeSurvey, while maintaining a consistent presentation of chat interactions. This standardization minimizes methodological artifacts caused by heterogeneous implementations and ensures comparability across studies. Finally, Simple Chat accommodates both text and vision models, with the option to disable vision functionality if desired. It can be configured via query parameters, as shown in Table~\ref{tab:params}, and system messages can be set through the admin interface. The admin interface starts with a login interface and it showcases the active projects in the lab, allowing for configuration of different conditions, each with their unique settings, such as system prompts. Conversations can be inspected during active studies, filtered by model, participant ID, experiment ID, and conversation ID, and downloaded.

\FigureTitle{table}{tab:params}{Required Query Parameters for Configuring Simple Chat}
\FigureBlock{51.4pt}{6.9pt}{\hbox{\TableSetup
  \begin{tabular}{p{60.2pt}>{\raggedright\arraybackslash}p{387pt}}
    \toprule
    \HeadRow\centering Parameter & \centering\arraybackslash Description \\
    \midrule
    \Row pid            & Project ID assigned by administrators upon project registration. \\
    \Row experiment\_id  & Identifier for the experiment, defined by the researcher. \\
    \Row participant\_id & Unique participant identifier, provided by the recruitment or experiment platform. \\
    \Row model          & Parameter specifying which LLM is used. \\
    \bottomrule
  \end{tabular}}}
\Note{Additional query parameters can be added as required.}
\blankline

\subsection{Performance}
\label{sec:performance}

To quantify the latency that participants experience, we benchmarked Simple Chat with GPT-5.1 (\code{gpt-5.1-2025-11-13}) via OpenAI and via Azure OpenAI (Sweden Central region), and with DeepSeek-V4.1-Flash via Together AI, each with streaming on and off and with 1, 5, or 20 concurrent chat sessions. Automated browsers, desktop Chrome and an emulated iPhone, opened the chat embedded in a web page, as in Qualtrics or oTree, and recorded when each reply appeared on screen; as a baseline, we called the providers' APIs directly (Figure~\ref{fig:latency}). The first characters of a reply appeared after a median of 0.55 to 2.04~s with streaming and 1.29 to 3.09~s without. Across model, concurrency, and device combinations, the 95th percentile of this on-screen time ranged from 0.67 to 20.30~s with streaming and 3.52 to 16.95~s without; streaming did not shorten this upper-tail wait in every condition. Compared with the direct calls, Simple Chat delayed the median by at most 0.6~s for GPT-5.1 and 2.2~s for DeepSeek-V4.1-Flash, whose response times varied more. Latency depended mainly on the model and provider, not on the number of concurrent sessions or the device, and no request failed. Because the interface reveals replies at 20~ms per character, complete replies took a median of 12.5 to 16.1~s to appear, mostly due to this display animation. The 95th percentile for complete replies ranged from 35.80 to 52.59~s with streaming and 35.97 to 42.50~s without. Researchers can shorten or remove the display animation in the frontend code (see \hyperref[sec:customizing]{Customizing Simple Chat}), so that complete replies appear as soon as the provider returns them.

\par\noindent\begin{minipage}{\linewidth}
\FigureTitle{figure}{fig:latency}{Time Until Participants See the LLM Reply in Simple Chat Compared With Direct API Calls}
\FigureBlock{0pt}{4pt}{\hbox{\includegraphics{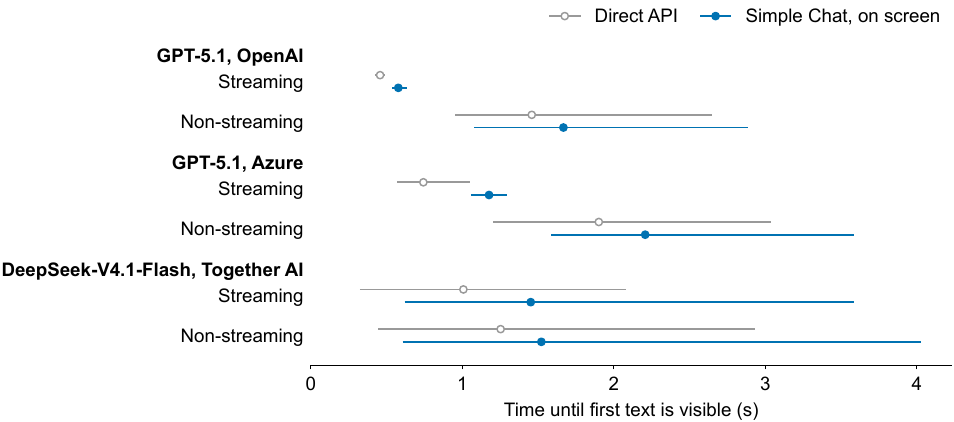}}}
\end{minipage}\par
\Note{Medians and interquartile ranges, pooled over desktop Chrome and an emulated iPhone and over 1, 5, and 20 concurrent sessions. Direct API is the time until the first text arrives when the provider is called directly: the first token with streaming, the complete response without.}

\subsection{Customization Features}
\label{sec:customization}

Simple Chat offers powerful customization features that enable researchers to tailor LLM interactions to the specific needs of their studies. One such feature is the ability to define custom system messages---special instructions that guide the LLM's behavior as a conversational assistant. In more advanced use cases, it may be necessary to generate a unique system message in real time for each participant and launch a Simple Chat window with that instruction. To achieve this, the system message can be sent during the experiment to a specific endpoint, which returns a system message ID. Including this ID as a query parameter (\codeV{system_message_id}) ensures that the customized instruction is applied.

Simple Chat also lets researchers specify that the assistant should initiate the conversation. If the \codeV{assistant_first} attribute is set to \code{True}, the LLM initiates the conversation as soon as the chat window opens. This is especially useful in studies where the LLM is expected to guide participants proactively from the start.

For studies involving multimodal interaction, Simple Chat supports vision-language models. In these cases, the interface can display a camera icon next to the send button, enabling participants to upload images directly within the conversation. This functionality can also be disabled with a query parameter, which gives researchers full control over whether image input is permitted in the respective study.

Finally, Simple Chat accommodates situations where the experiment might interact with the chatbot without the need of a full conversation. With the direct call feature, researchers can send a specific instruction to the model via an HTTP request and receive the response directly as a string. This makes it possible to integrate both conversational and non-conversational LLM interactions seamlessly into online experiments, and share context between both types of calls.

\subsection{Setup and Deployment}
\label{sec:deployment}

This section describes how researchers deploy the open-source version of Simple Chat on their own server. Researchers who use the hosted service can skip it. Simple Chat consists of four services that run in containers: the frontend, the backend, the dashboard, and the database (see Figure~\ref{fig:technical-design} and Table~\ref{tab:services}). An object storage service is needed only if participants upload images. All components communicate through HTTP APIs. The technologies used to implement each service are listed in Appendix~B, and the database schema is shown in Appendix~A. The controls that restrict access to a deployment and limit its costs are described in \hyperref[sec:security]{Security and Cost Containment}.

\FigureTitle{figure}{fig:technical-design}{Overview of the Technical Design}
\FigureBlock{218.2pt}{1pt}{\hbox{\hspace*{1.5pt}\includegraphics[width=468pt,height=187.6pt]{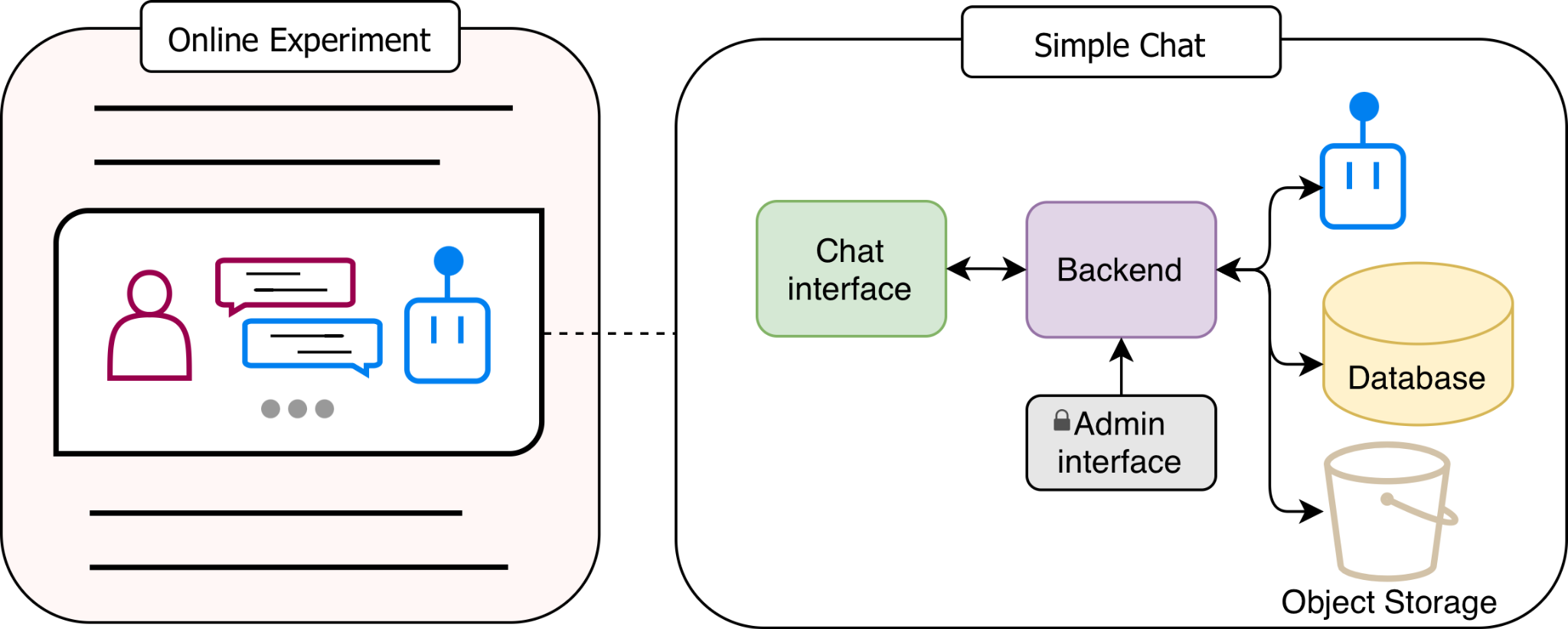}}}

\FigureTitle{table}{tab:services}{Services of Simple Chat}
\FigureBlock{98pt}{6.9pt}{\hbox{\TableSetup
  \begin{tabular}{p{70pt}>{\raggedright\arraybackslash}p{376pt}}
    \toprule
    \HeadRow\centering Service & \centering\arraybackslash Description \\
    \midrule
    \Row[12pt]Frontend       & Chat interface that participants see, embedded in the survey as an iframe. \\
    \Row[12pt]Backend        & Communicates with the LLM providers, manages projects and conversations, and provides the API used by the frontend, the dashboard, and the survey platform. \\
    \Row[12pt]Dashboard      & Administrative interface in which researchers configure projects and inspect and download conversations. \\
    \Row[12pt]Database       & Stores conversation transcripts, participant interaction data, and project settings. \\
    \Row[12pt]Object storage & Stores images that participants upload during chats. Optional; needed only when image input is enabled. \\
    \bottomrule
  \end{tabular}}}

\subsubsection{Running Simple Chat on a Single Server}
\label{sec:docker-compose}

The open-source repository includes a Docker Compose file that builds and starts all four services on a single server. Deployment involves four steps. First, researchers assign three domain names to the server, one each for the chat, the dashboard, and the backend. Second, they clone the repository, copy the example configuration file \code{.env.example} to \code{.env}, and enter their LLM provider API key, the dashboard credentials, and the public addresses of the chat and the backend. Third, they build and start all services with a single command:

\par\noindent\hspace*{0.5in}\code{docker compose --profile prod up --build --detach}\par

\noindent Fourth, they configure a reverse proxy, a web server that forwards requests to the three services, so that all three addresses are served over HTTPS and data transfer is encrypted. The repository's README documents these steps, provides an example reverse-proxy configuration, and covers updates, database backups, and local testing. Box~\ref{box:hosting} compares common hosting options.

\begin{InfoBox}{box:hosting}{Deploying on a Hosting Provider}
\noindent Simple Chat runs on any server with Docker and the Docker Compose plugin. The choice of host mainly determines how quickly a server is available, who maintains it, and where participant data are stored.

\smallskip\noindent\textbf{Commercial hosting providers.} A virtual server can be created within minutes through a web interface and is billed by usage. Many providers operate data centers in several regions, including the European Union, so researchers can choose where participant data are stored.

\smallskip\noindent\textbf{Institutional virtual machines.} Many universities and research institutes provide virtual machines through their IT department. Data then remain on institutional infrastructure, which can simplify data protection and ethics review. In exchange, provisioning usually takes longer, and the IT department may need to assign domain names, issue TLS certificates, and open ports.

\smallskip\noindent\textbf{Platform services.} Platforms such as Railway build and run containers without a server to administer. Railway does not run Docker Compose files directly: each service in the Compose file becomes a separate Railway service built from its folder in the repository, the database can be replaced by Railway's managed MongoDB, and Railway generates a public domain for each service that participants or researchers open (\url{https://railway.com/guides/docker-compose}).

\smallskip\noindent With a commercial provider, researchers are responsible for operating system updates and backups, and data protection law such as the GDPR may require a data processing agreement with the provider.

\smallskip\noindent As a starting point, we recommend a server with two virtual CPUs, 4~GB of memory, and 25~GB of disk space. Most of this memory is needed to build the containers. Once running, Simple Chat mainly forwards messages to the LLM provider and needs few resources.
\end{InfoBox}

\subsubsection{Customizing Simple Chat}
\label{sec:customizing}

Researchers who want to change Simple Chat itself can modify the open-source code and rebuild the containers with the same Docker Compose file. The repository's contributing guide describes how to set up a development environment for this purpose.
\blankline

\subsection{Security and Cost Containment}
\label{sec:security}

Because participants open the chat through a URL, anyone who obtains this URL can use the chat outside the intended study flow, which may corrupt the data and raise API costs. Deactivating a project prevents such use, but only for all participants at once. For finer control, researchers can configure four types of safeguards for each project in the dashboard. Simple Chat leaves them disabled by default because suitable values depend on the study, and some limits, such as the number of turns, shape the experimental manipulation itself. Researchers therefore set them as part of the study design.

First, usage limits restrict how much each participant and the project as a whole can consume. Researchers can limit the number of conversations per participant, the number of turns per conversation, and the number of tokens per conversation and per project. The token limits are enforced before each model call: Simple Chat reserves room for the response in advance, so that neither a single response nor simultaneous conversations can exceed them. Once a limit is reached, the chat closes and the participant sees a brief closing message.

Second, an allowlist admits only known participants, such as the Prolific IDs of the recruited sample, and a rate limit prevents participants from sending messages in rapid succession. Because participant IDs are passed in the URL, the allowlist cannot stop someone who knows a valid ID; in this case, the project's token limit still caps the total cost.

Third, researchers can specify the websites on which the chat may be embedded, such as their survey platform, and browsers then refuse to display it elsewhere. This restriction does not apply when the chat URL is opened directly, so it should be combined with the allowlist and the usage limits.

Fourth, the dashboard reports token usage and estimated costs per project, per model, and per conversation, which allows researchers to monitor expenses during data collection. Configuration details for all safeguards are documented in the repository.

\subsection{Comparison With Existing Tools}
\label{sec:comparison}

Table~\ref{tab:features} compares Simple Chat with the tools introduced in the section Existing Methods and Challenges of Integrating LLMs Into Online Experiments. It covers the properties that typically decide which tool fits a study: the platforms and model providers a tool supports, whether it streams responses and accepts images, how conversations are logged and exported, how sessions and costs are controlled, what it takes to deploy the tool, how it can be extended, and its license. Each entry reflects the tool's public code or documentation as of November 6, 2026. The table compares documented features only. It does not compare response times or participant behavior across tools; doing so would require running the same study with each tool, which is beyond the scope of this tutorial.

\newsavebox{\featuretable}
\sbox{\featuretable}{\TableSetup\setlength{\tabcolsep}{4pt}\fontsize{8.5}{10}\selectfont
  \begin{tabular}{>{\raggedright\arraybackslash}p{64pt}*{6}{>{\raggedright\arraybackslash}p{58pt}}}
    \toprule
    \HeadRow Feature & Simple Chat & SCALE & oTree GPT & G4R & LEXI & Qualtrics template \\
    \midrule
    \Row Platforms & Any that embeds an iframe (e.g., Qualtrics, oTree) & oTree; standalone page & oTree & Qualtrics; browser tab & Standalone app & Qualtrics \\[4pt]
    \Row Providers & Any with an OpenAI-compatible API, incl.\ open-weights models & OpenAI & OpenAI & OpenAI & OpenAI & Via OpenRouter \\[4pt]
    \Row Streaming & Yes & No & No & -- & Yes & No \\[4pt]
    \Row Input & Text, images & Text & Text; voice in one template & Text & Text & Text \\[4pt]
    \Row Logging and export & CSV, JSON; dashboard & DynamoDB; scripts & oTree export & CSV & Excel & Qualtrics data \\[4pt]
    \Row Authentication and sessions & IDs in URL; ID allowlist; origin check; dashboard login & Origin check; ID allowlist & oTree sessions & Researcher accounts & Admin login; sign-up & Qualtrics sessions \\[4pt]
    \Row Cost controls & Token, turn, chat and rate limits; usage view & Token, message and chat limits & -- & Message limit & Message, chat and participant limits & Token and round limits \\[4pt]
    \Row Deployment & Self-hosted or hosted service & AWS Lambda & oTree server & Hosted service & Node.js; MongoDB & Qualtrics import \\[4pt]
    \Row Extensibility & Open source & Open source & Open source & Settings form & Open source & Survey JavaScript \\[4pt]
    \Row License & CC BY 4.0 & MIT & MIT & Paper: CC BY 4.0; no public code & CC BY-NC 4.0 & -- \\[2pt]
    \bottomrule
  \end{tabular}}
\NewPageUnlessBoxFits{\featuretable}
\FigureTitle{table}{tab:features}{Features of Simple Chat and Existing Tools for Integrating LLMs Into Online Experiments}
\BoxBlock{\featuretable}
\Note{Entries are based on the public code or documentation of each tool, checked on November 6, 2026: SCALE \parencite{valet2025scale}, oTree GPT \parencite{mckenna2023otree}, G4R \parencite{kim2025capture}, LEXI \parencite{laban2024lexi}, and the Qualtrics template of \textcite{costello2024durably}. -- = not documented. AWS = Amazon Web Services.}

The table shows that the tools trade breadth against ease of setup. The Qualtrics template and oTree GPT run inside their host platform and need no additional server, but each serves only that platform. G4R needs no installation because it is a hosted service, but it supports only OpenAI's models, and we found no public source code for it. SCALE and LEXI are open source but need their own infrastructure: an AWS account for SCALE, and a Node.js server with a MongoDB database for LEXI. Simple Chat can be embedded in any platform that displays an iframe, works with several providers, including open-weights models, and streams responses. Researchers who run their own instance deploy and maintain it themselves (see Setup and Deployment). Those who use the hosted service, Simple Chat as a Service, need no installation and only embed an iframe in their survey. Per-project limits on tokens, turns, conversations, and request rates contain the cost of misuse (see \hyperref[sec:security]{Security and Cost Containment}).
\blankline

\section{Quick Start}
\label{sec:quick-start}

The fastest way to a working chat is Simple Chat as a Service, which needs no installation. The following five steps lead from registration to the export of a test conversation. Researchers who host Simple Chat themselves first deploy it as described in Running Simple Chat on a Single Server. The steps in their own dashboard are similar, except that they sign in with the credentials from the \code{.env} file, where the API key is also set.

\textbf{Step 1: Register.} Open the dashboard at \url{https://simple-chat-dashboard.chm.mpib-berlin.mpg.de}, click ``Register,'' and enter an email address and a password. Click the activation link sent by email within 48 hours, then sign in.

\textbf{Step 2: Create a project.} In the sidebar, click ``Create New Project'' and enter a project ID of lowercase letters, numbers, and underscores (e.g., \code{my_study}), the API key from the LLM provider, the name of the model as given by the provider, and a system message (Figure~\ref{fig:create-project}). The API endpoint is needed only for providers other than OpenAI. Then click ``Create the project.'' Model usage is billed to the provider account of the API key. A project with an API key is active as soon as it is created.

\FigureTitle{figure}{fig:create-project}{Creating a Project in the Dashboard}
\FigureBlock{318pt}{8.3pt}{\hbox{\hspace*{1.5pt}\setlength{\fboxsep}{0pt}\setlength{\fboxrule}{0.75pt}\fbox{\includegraphics[width=420pt]{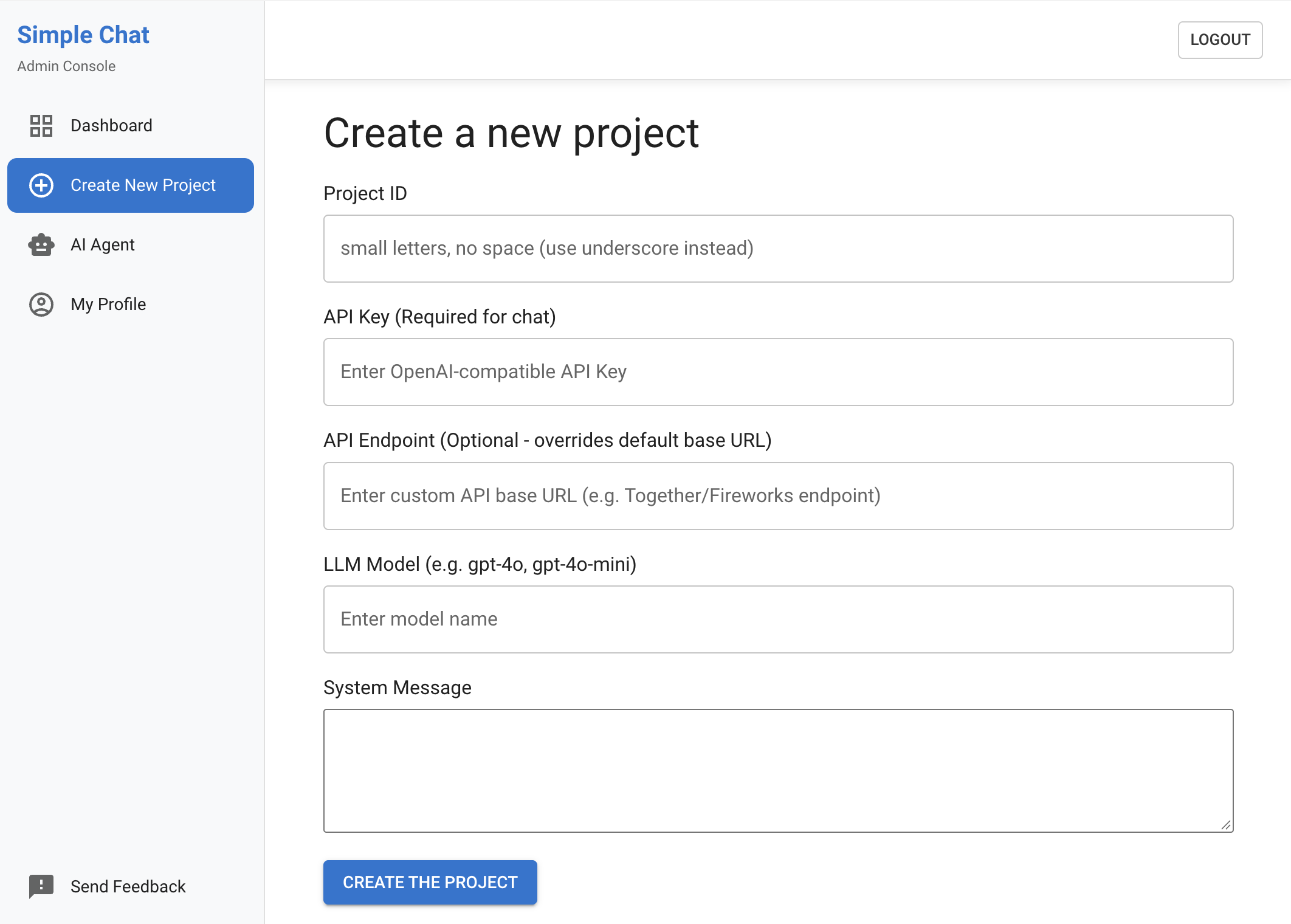}}}}

\textbf{Step 3: Embed the chat.} In Qualtrics, add a Text/Graphic question, switch its editor to HTML view, and paste the following code, with the URL on one line:
\CodeFile[style=html]{code/quickstart-iframe.html}
Here, \code{pid} is the project ID, the Qualtrics piped text in \code{participant_id} passes each participant's response ID, and \code{experiment_id} labels the condition (Table~\ref{tab:params}). In oTree, the same code goes into the HTML template of the page, with \code{{{ participant.code }}} as the participant ID.

\textbf{Step 4: Test a conversation.} Preview the survey, send a few messages, and check in the dashboard under Project Settings $\rightarrow$ Conversations that the conversation was recorded with the expected participant ID.

\textbf{Step 5: Export the data.} On the same page, click ``Download All'' to save all conversations of the project as JSON, or filter the list first and download only the filtered conversations. When data collection ends, click ``Deactivate Project'' under System Status so that the chat no longer accepts participants.

The two case studies below build on these steps and describe only what each study adds.
\blankline

\section{Case Studies}
\label{sec:case-studies}

But what might a practical application of Simple Chat look like? To demonstrate its potential, we present two case studies that illustrate how existing online experiments on human--LLM interactions can be re-implemented with Simple Chat.

\subsection{Conversations With LLMs as an Intervention to Reduce Belief in Conspiracy Theories}
\label{sec:qualtrics}

\textcite{costello2024durably} conducted a series of online survey experiments using Qualtrics to evaluate the effectiveness of dialogues with LLMs in reducing belief in conspiracy theories. As part of the experiments, participants were asked to report a conspiracy theory they personally believed in. They were then randomly allocated to either an intervention condition or a control condition. Participants in the intervention condition engaged in a three-round conversation with an LLM, specifically GPT-4 Turbo, which had been prompted to address their specific conspiracy theory directly and to persuade them that the belief was unfounded. Conversely, those placed into the control condition also interacted with GPT-4 Turbo but discussed an unrelated topic. The intervention resulted in a substantial reduction in conspiracy beliefs, with this effect persisting even two months later. These findings highlight the potential of LLM-based conversations as a powerful tool for countering belief in conspiracy theories.

To re-implement the setup of \textcite{costello2024durably} with Simple Chat, create a project as in Step 2 of the Quick Start with the project ID \code{conspiracy_with_ai}. Because the debunking prompt is personalized for each participant and set during the experiment, as described below, the system message entered in the dashboard only serves as a fallback for chats opened without a valid \code{system_message_id}.

Notably, \textcite{costello2024durably} had LLMs create summaries of participants' conspiracy theory descriptions. Prior to the conversation with the LLM, participants were first asked whether they believed in any conspiracy theory and, if so, to specify which one. They were also asked to provide details about why they found the respective theory compelling, such as particular pieces of evidence. This information was then passed to an LLM, which generated a real-time summary of each participant's free-text description. Participants were subsequently asked to indicate how strongly they believed this LLM-generated summary. This served as a pretreatment measure of conspiracy belief. Simple Chat provides an endpoint that calls an LLM and returns the response as a string, which simplifies this step of the experiment. The collected input can be sent to \codeV{[POST] /llm/call} with \code{project_id}, \code{requested_by}, \code{model}, and \code{chat}. It is saved as a conversation in the Simple Chat database in the \codeV{conspiracy_with_ai} project.

In the next phase of the experiment, all participants took part in a three-round dialogue with an LLM. Participants in the intervention condition discussed their conspiracy theory in a debunking conversation with the LLM, whereas those in the control condition engaged in a discussion on an unrelated topic. The system message employed by \textcite{costello2024durably} for the LLM-powered debunking dialogue contained details about the context, the goal of refuting the conspiracy belief, and guidelines for how the model should behave. These elements remained consistent across all model calls. In addition, the system message included participant-specific information: the participant's particular conspiracy theory, their stated reasons for believing it, and their level of belief in the theory. The participant-specific information was dynamically inserted into the instructions for each individual participant. Before directing the user to the screen with the iframe, the Qualtrics backend must first prepare and send the personalized prompt via a call to \codeV{[POST] /project/custom_system_message}. This request returns a \codeV{system_message_id}, which must then be included as a query parameter like \codeV{&system_message_id=...} in the iframe URL, as shown in Figure~\ref{fig:qualtrics-iframe}. When the iframe is initialized, Simple Chat uses the provided \codeV{system_message_id} to automatically set the personalized prompt at the start of the conversation, ensuring it is applied before the interaction begins.

\FigureTitle{figure}{fig:qualtrics-iframe}{Example iframe for the First Round in the Qualtrics Experiment}
\FigureBlock{133.4pt}{8.6pt}{\hbox{\hspace*{1.4pt}\CodeBox{403.7pt}{123.6pt}{
\begin{lstlisting}[style=html]
<iframe width="90%" height="500"
src="https://YOUR_SIMPLE_CHAT_FRONTEND_URL/pid=conspiracy_with_ai&
model=gpt4o&
experiment_id=human_ai_interaction_experiment&
participant_id=${e://Field/ResponseID}&
upload_image=false&
session_id=${e://Field/SessionID}"
></iframe>
\end{lstlisting}}}}
\Note{\codeV{YOUR_SIMPLE_CHAT_FRONTEND_URL} is a placeholder for the deployed version of Simple Chat.}
\blankline

The iframe is inserted into a Qualtrics question in the HTML view, as in Step 3 of the Quick Start. In this study, the \code{src} attribute of the iframe includes four parameters: \codeV{pid}, \code{model}, \code{experiment_id}, and \code{participant_id}. Details on these parameters can be found in Table~\ref{tab:iframe-params}.

\FigureTitle{table}{tab:iframe-params}{Parameters of the \code{src} Attribute of the iframe}
\FigureBlock{117.9pt}{9.1pt}{\hbox{\TableSetup
  \begin{tabular}{>{\ttfamily}p{93.9pt}>{\raggedright\arraybackslash}p{352.3pt}}
    \toprule
    \HeadRow\normalfont Parameter & Description \\
    \midrule
    \Row[13pt]pid            & The parameter is set to \texttt{conspiracy\_with\_ai}. \\
    \Row[12pt]model          & The parameter specifies one of the models defined in the Simple Chat configuration. When models are used as experimental conditions, this value can be defined as a variable and passed through the Qualtrics survey flow\fix{}{.} \\
    \Row[12pt]experiment\_id  & The parameter indicates the treatment condition assigned to the participant. \\
    \Row[12pt]participant\_id & The parameter serves as a unique identifier for the task and makes it possible to match user data from Qualtrics or other platforms with the data recorded in Simple Chat. \\
    \bottomrule
  \end{tabular}}}

During or after the data collection phase, the conversations are exported from the Simple Chat dashboard as in Step 5 of the Quick Start, or directly from the MongoDB database. Figure~\ref{fig:dashboard} shows the dashboard with the conversations of the \code{conspiracy_with_ai} project.

\textcite{costello2024durably} used survey blocks not only to display LLM messages and participant input fields but also to contain the JavaScript logic needed to query the LLMs. With Simple Chat, this logic is moved to a dedicated server and embedded in the experiment as an iframe, which abstracts the underlying complexity. It is important to note that the links used in iframes can be accessed independently by users, which may result in data corruption and increased API costs. This is a known issue from this approach, which is also present in SCALE
\fix{(Valet \& Werner, \citeyear{valet2025scale})}{\parencite{valet2025scale}}.
To mitigate these risks, each project can be deactivated in the dashboard when it is not in use, and finer-grained controls limit who can use the chat and how much (see \hyperref[sec:security]{Security and Cost Containment}).

\FigureTitle{figure}{fig:dashboard}{Simple Chat Dashboard}
\FigureBlock{377.2pt}{1pt}{\hbox{\hspace*{1.5pt}\includegraphics[width=378pt,height=349.3pt]{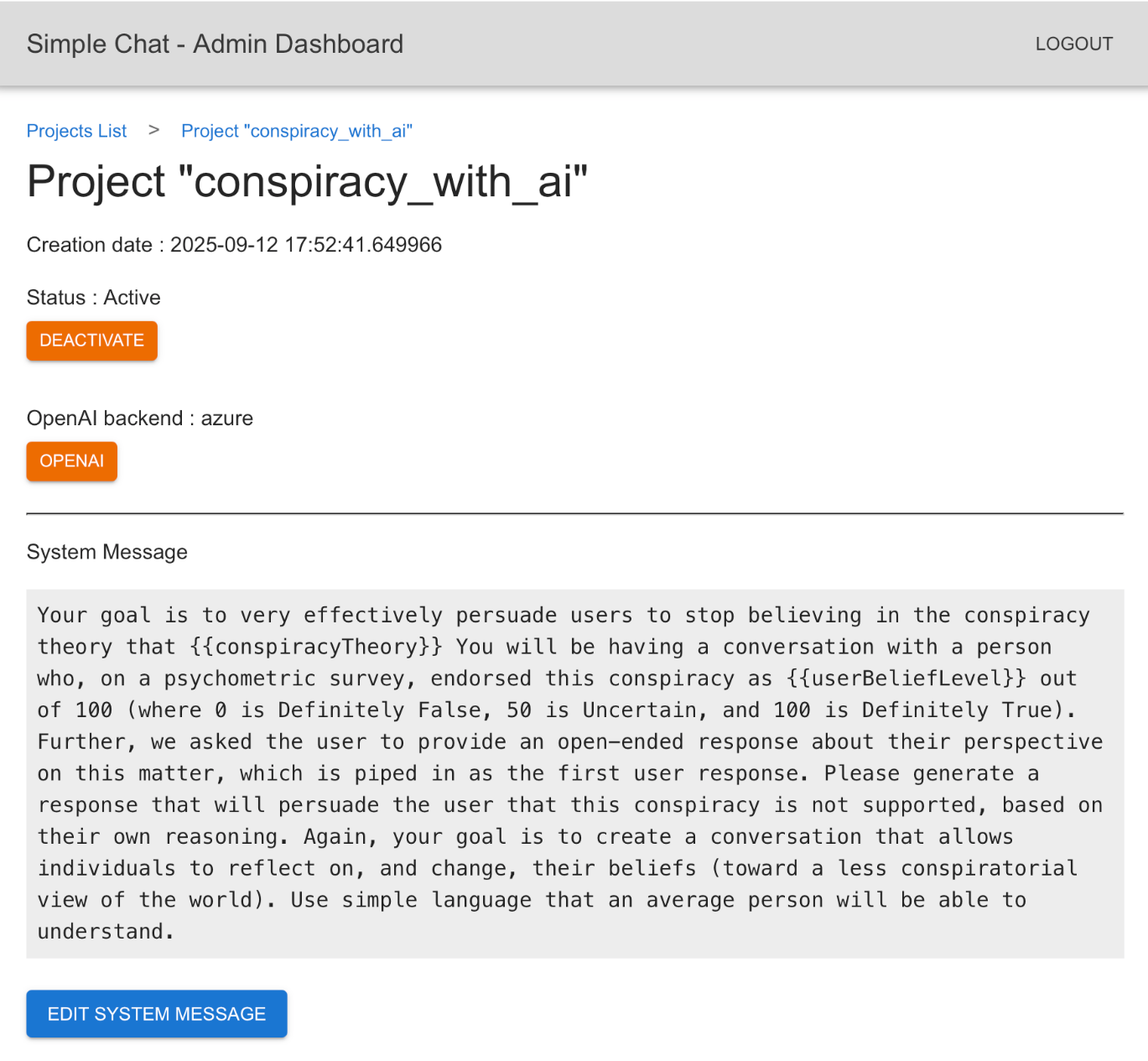}}}

\textcite{costello2024durably} provide a Qualtrics template that can be used to replicate their studies or to conduct other types of human--LLM interaction experiments. Following the steps outlined above, we developed a Simple Chat version of this template to demonstrate what a human--LLM online experiment built with Simple Chat might look like. This version of the template is available in our GitHub repository (\url{https://github.com/center-for-humans-and-machines/simple-chat}).

\subsection{Conversations With LLMs as an Intervention to Address Political Polarization}
\label{sec:otree}

\textcite{walter2025using} examined the potential of LLMs to mitigate political polarization through an online survey experiment implemented with oTree. Participants were randomized into either a treatment condition or a control condition, each involving a six-minute conversation with an LLM, namely GPT-4o, about the highly polarized issue of U.S. support for Ukraine. For the treatment condition, GPT-4o was prompted to encourage participants toward a centrist position, whereas for the control condition it was configured to validate participants' initial views and reinforce their confidence in them. The depolarization version of GPT-4o successfully reduced ideological polarization, and this effect was still evident one month later, albeit weaker. However, the effect was limited to certain participants, and the treatment had only a modest impact on affective polarization. These findings suggest that LLMs hold some promise as tools for depolarization but also underscore the need for caution, given the risks of exploiting their persuasive power.

The first step in re-implementing this study with Simple Chat is to create a project as in Step 2 of the Quick Start, using the dashboard as shown in
Figure~\fix{\ref{fig:qualtrics-iframe}}{\ref{fig:create-project}}.
In this case, the project ID may be set to \code{mitigate_political_polarisation}. As in the Qualtrics case study, the project's system message only serves as a fallback, because the two system prompts of the study are uploaded when the oTree server starts, as described below.
\blankline

As in the studies by \textcite{costello2024durably}, the key difference between the treatment and control conditions in
\fix{Werner's (\citeyear{walter2025using})}{\citeauthor{walter2025using}'s (\citeyear{walter2025using})}
study lies in the system prompt. As described earlier, distinct prompts were used to create a ``depolarization'' version of the LLM for the treatment condition and a ``confirming'' version for the control condition. The ``depolarization'' prompt supplies the LLM with contextual information about the depolarization task, specifies the intended content of the messages, provides stylistic guidelines, and includes a list of arguments that the LLM should use to encourage participants to adopt more moderate views on the polarized topic of U.S. support for Ukraine. Both prompts were stored as separate text files due to their considerable length and because they do not contain placeholder elements, unlike the prompts used by \textcite{costello2024durably}. They are automatically uploaded to Simple Chat upon initialization of the oTree server. The system prompts are stored by sending a request to the Simple Chat backend server (\codeV{[POST]/project/custom_system_message}). Participants are randomly assigned to either the ``confirming'' or ``depolarizing'' condition, and the appropriate system message is activated by including the corresponding \codeV{system_message_id}, obtained in the previous step, as a query parameter in the URL used to render the iframe within the experiment.

An iframe within the oTree HTML template, illustrated in Figure~\ref{fig:otree-iframe}, provides the environment for all LLM chats. Besides the dashboard export in Step 5 of the Quick Start, each chat can be retrieved by calling the chat-downloading endpoint from the oTree Python code, as shown in Figure~\ref{fig:otree-download}.

\FigureTitle{figure}{fig:otree-iframe}{Integration as iframe in the oTree Experiment}
\FigureBlock{121.3pt}{7.7pt}{\hbox{\hspace*{1.8pt}\CodeBox{413pt}{113pt}{
\begin{lstlisting}[style=html,basicstyle=\linespread{1}\codefont\fontsize{8}{10.3}\selectfont\color{codetext}]
<iframe
  width="90%"
  src="https://YOUR_SIMPLE_CHAT_FRONTEND_URL/?pid=mitigate_political_polarisation&
model={{ llm-model }}&
system_message_id={{ custom_system_message_id }}&
experiment_id={{ experiment_id }}&
participant_id={{ participant_id }}&
upload_image=false"
></iframe>
\end{lstlisting}}}}
\Note{\codeV{YOUR_SIMPLE_CHAT_FRONTEND_URL} is a placeholder for the deployed version of Simple Chat.}
\blankline

\FigureTitle{figure}{fig:otree-download}{Downloading Chat Data from Simple Chat via oTree}
\FigureBlock{235.3pt}{8.5pt}{\hbox{\hspace*{1.7pt}\CodeBox{355.1pt}{225.5pt}{
\begin{lstlisting}[style=python,basicstyle=\linespread{1}\codefont\fontsize{10.4}{13.2}\selectfont\color{codetext}]
import requests

url = f"""YOUR_SIMPLE_CHAT_BACKEND_URL/download/chat/
mitigate_political_polarisation/
{experiment_id}/{participant_id}
""".replace("\n", "")

try:
    response = requests.get(url)
    response.raise_for_status()
    data = response.json()
    print(data)

except requests.exceptions.RequestException as e:
    print("Error during request:", e)
\end{lstlisting}}}}
\Note{\codeV{YOUR_SIMPLE_CHAT_BACKEND_URL} is a placeholder for the deployed version of Simple Chat.}
\blankline

While \textcite{walter2025using} does not provide the source code needed to replicate the experiment, the underlying SCALE framework (Serverless Chat Architecture for LLM Experiments;
\fix{Valet \& Werner, \citeyear{valet2025scale}}{\cite{valet2025scale}})
on which it is built is publicly available. SCALE provides a chat interface delivered through a Software-as-a-Service setup and operates via an Amazon Web Services (AWS) Lambda function that communicates with the oTree frontend. In line with the steps described above, we created an updated version of SCALE with Simple Chat included, which can be retrieved from our GitHub repository (\url{https://github.com/center-for-humans-and-machines/simple-chat-otree-walter-2025}).

\section{Discussion}
\label{sec:discussion}

Simple Chat simplifies and optimizes the integration of LLMs into online experiments in three key ways. First, it removes the need for custom-coded API connections by offering a ready-to-use, embeddable chat interface that is compatible with widely used survey platforms. Second, it standardizes the experimental interface across platforms and studies, reducing variability in chat presentation and minimizing methodological artifacts that may arise from inconsistent designs. Third, it supports streaming completions, which display responses incrementally as tokens are generated and thus shorten the wait for visible text.

Table~\ref{tab:features} sets these properties against those of existing tools. Among the tools compared there, Simple Chat is the only one that can be embedded in any survey platform and also works with several model providers, including open-weights models. Whether its standardized interface yields participant behavior equivalent to that in other setups remains to be tested.

By lowering technical complexity and standardizing the experimental interface, Simple Chat can make human--LLM research more accessible and consistent across studies. With the hosted service, researchers add an LLM to an online experiment by embedding an iframe, without setting up or maintaining a server. Researchers who deploy or adapt Simple Chat themselves can reuse a single deployment across studies and survey platforms. Beyond making online experiments with LLMs more accessible for researchers, Simple Chat fosters a more unified research ecosystem in which results can be compared, replicated, and extended more effectively. In this way, Simple Chat strengthens the foundations for rigorous study of human--LLM interactions and accelerates the cumulative advancement of knowledge in this rapidly evolving field.

\subsection{Ethical Considerations}
\label{sec:ethics}

Naturally, the question arises as to whether researchers are obliged to obtain informed consent whenever participants are exposed to LLMs. Informed consent should state clearly that participants will be interacting with an LLM rather than a human, outline potential risks, and specify how data will be collected, stored, and used, thereby enabling participants to make an informed decision about their involvement in the study. \textcite{bail2024can} cautions, however, that this kind of disclosure may limit the usefulness of LLMs for simulating human behavior, as participants' reactions could be driven by their attitudes toward AI rather than by the interaction itself. In any case, when studies on human--LLM interaction involve deception or withhold information about the use of an LLM, researchers must obtain explicit approval from the relevant ethics committee.

Responsible use of Simple Chat also requires a strong commitment to data privacy. As \textcite{behrend2025participant} point out, researchers should carefully consider whether participants may share sensitive information while interacting with the LLM. More broadly, as with conventional studies, data collection should be limited strictly to what is necessary to achieve the research objectives, and all records must be stored securely in accordance with applicable privacy regulations (e.g., GDPR). This applies not only to questionnaire responses but also to any information generated during interactions with the LLM.

Moreover, it should be taken into account that LLMs can produce inappropriate and potentially even harmful outputs \parencite[e.g.,][]{wen2023unveiling}. Simple Chat relies solely on safety mechanisms implemented by the LLM providers and does not include independent safety monitoring or filtering features. Therefore, researchers should exercise caution and consider implementing additional safeguards, especially in studies involving vulnerable populations or sensitive topics. Such safeguards can run alongside Simple Chat, for example, a hybrid monitoring system that keeps clinicians in the loop \parencite{yun2026hybrid}.

In practice, many of these safeguards depend on how Simple Chat is configured and deployed. Table~\ref{tab:governance} summarizes what the software stores and where, and which settings and decisions remain with the research team. Free-text input warrants particular care: Simple Chat stores each message exactly as typed, so any personal information that participants disclose ends up in the database, in data exports, and with the LLM provider. The same holds for the query parameters of the chat URL, which are saved with each conversation and included in the CSV export; the URL should therefore carry only pseudonymous identifiers (see Table~\ref{tab:params}). Where these data reside depends on the chosen deployment (see \hyperref[sec:deployment]{Setup and Deployment}). These details also belong in the consent form discussed above. For studies that disclose the use of an LLM, the following wording may serve as a starting point, with the bracketed parts adapted to the study:
\begin{quote}
In this study, you will chat with an artificial intelligence (AI) system, not with a human. The AI's answers are generated automatically and may be inaccurate or inappropriate. Your messages [and any images you upload] are sent to [provider] to generate the answers and are stored, together with a pseudonymous participant ID, on servers operated by [institution] until [date]. [Provider] may keep your messages for up to [period]. Please do not enter names, contact details, or other information that could identify you or other people. You can withdraw from the study at any time.
\end{quote}

\FigureTitle{table}{tab:governance}{Data-Governance Checklist for Studies Using Simple Chat}
\nopagebreak\vspace{6pt}\nopagebreak
{\TableSetup\noindent
  \begin{tabular}[b]{>{\raggedright\arraybackslash}p{62pt}>{\raggedright\arraybackslash}p{172pt}>{\raggedright\arraybackslash}p{203pt}}
    \toprule
    \HeadRow Item & Simple Chat behavior & What researchers configure or decide \\
    \midrule
    \Row Stored data & MongoDB stores each conversation (messages with timestamps, system message, participant, experiment, and conversation IDs), all URL query parameters, token usage per participant, and direct calls. The backend log records message text. & Host MongoDB and object storage on infrastructure covered by the institution's data-protection rules or in a suitable region of a managed service (see \hyperref[sec:deployment]{Setup and Deployment}). Limit who can read the server logs. \\
    \addlinespace[2pt]
    \Row Participant IDs & The \code{participant_id} value is taken from the URL as is and appears in the dashboard and all exports. & Pass only pseudonymous IDs from the survey platform; never put names, email addresses, or other personal data into any query parameter. Keep the key linking IDs to identities separately. \\
    \addlinespace[2pt]
    \Row Images & The upload button appears only for the vision models listed in the frontend code; \code{upload_image=false} hides it. Images are saved in object storage and passed to the provider as a link, which expires after 1 hour; the image is kept. & Set \code{upload_image=false} unless the design requires images. Otherwise use a private bucket and ask participants not to upload pictures of people or documents. \\
    \addlinespace[2pt]
    \Row LLM provider & Messages, the system message, and image links are sent to the LLM provider configured for the project (e.g., OpenAI, Azure OpenAI, or Together AI). & Conclude a data-processing agreement. Prefer zero data retention, no training on inputs, and a permitted processing region. For sensitive topics, use an open-weights model hosted on institutional clusters, which Simple Chat can call through an OpenAI-compatible API. \\
    \addlinespace[2pt]
    \Row Observability & Sentry and Langfuse are off by default. If enabled, Langfuse receives prompts, responses, and participant IDs; its default host is Langfuse's cloud service. & Enable them only with a self-hosted instance or one covered by a data-processing agreement, and mention them in the consent form. \\
    \addlinespace[2pt]
    \Row Retention and deletion & Nothing is deleted automatically, and the dashboard has no delete function. Only rate-limit records keyed by IP address expire, after 24 hours. & Set a retention period in the data-management plan. After export, delete the records and images directly in the database and the bucket, including backups. \\
    \addlinespace[2pt]
    \Row Participant requests & The dashboard filters conversations by participant ID and downloads them as JSON; the backend exports CSV files per conversation and experiment. & Answer access requests with a filtered export. For erasure requests, delete every record with that \code{participant_id} and the images linked in its messages. \\
    \addlinespace[2pt]
    \Row Consent & Simple Chat shows no information or consent page. & Obtain consent in the survey before the chat, adapting the sample wording above. \\
    \addlinespace[2pt]
    \Row Content safety & No built-in content filtering or moderation; Simple Chat relies on the provider's safety mechanisms. & Test the system message with adversarial inputs, review conversations in the dashboard during data collection, and prepare a debriefing with support contacts for sensitive topics. Monitor externally if needed \parencite{yun2026hybrid}. \\
    \addlinespace[2pt]
    \Row Access & The dashboard requires researchers to sign in; data exports by survey platforms require a project-specific key. & Grant dashboard accounts only to members of the research team, and do not disclose the project-specific key to participants. \\
    \bottomrule
  \end{tabular}\par}
\nopagebreak
\Note{Direct calls are the HTTP requests described in \hyperref[sec:customization]{Customization Features}.}
\blankline

\subsection{Limitations}
\label{sec:limitations}

Aside from these ethical considerations, researchers intending to use Simple Chat should also be mindful of several inherent limitations. One key concern relates to usability. Self-hosting Simple Chat involves the command line, Docker, and the configuration of domain names and HTTPS, so researchers who have not worked with these tools may need support from their institution's IT department. The hosted service removes this step but ties researchers to infrastructure operated by the Center for Humans and Machines. To support researchers who deploy Simple Chat themselves, Setup and Deployment describes the installation step by step, the Quick Start section guides them from project creation to data export, and the repository provides further documentation and a tutorial video.

The case studies show how the experimental procedures can be implemented with Simple Chat; we did not collect data with them. Simple Chat provides a consistent interface across survey platforms, which can support comparisons among studies that use it. Because interface design may affect participant behavior, whether the original findings would also hold with these implementations remains an empirical question.

The need for continuous maintenance presents another notable challenge. Although the current version provides a powerful tool for embedding LLMs into experiments, the rapidly evolving landscape of LLM technology demands continuous updates to keep pace with new features. Since the initial development of Simple Chat, advancements such as streaming chat completions, support for reasoning models, and multimodal functionality including image uploads have emerged. To address this, we aim to cultivate an open-source community around Simple Chat, encouraging contributions to help ensure its ongoing development and relevance.

We also did not measure whether Simple Chat affects participant engagement or attrition. Streaming and the familiar chat interface were designed to reduce frustration and dropout, but whether they do so in online experiments, compared with non-streamed responses or other interfaces, is a hypothesis for future work. Studies that use Simple Chat can test it by randomizing streaming and comparing abandonment rates between conditions.

Finally, API costs remain an important financial consideration for researchers. Simple Chat tracks token usage and estimated costs per project and lets researchers limit conversations, turns, tokens, and request rates (see \hyperref[sec:security]{Security and Cost Containment}). Because suitable limits depend on each study's design and budget, these safeguards are disabled by default; a project without configured limits is therefore protected only by deactivation and a fixed cap on the size of each request, so researchers should set the limits before data collection begins. Moreover, the cost estimates rely on configured list prices and estimated token counts, so they can deviate from the provider's invoice. For detailed cost monitoring, researchers can additionally use the dashboards offered by LLM service providers or open-source tools such as Langfuse.

\subsection{Conclusion}
\label{sec:conclusion}

In sum, Simple Chat offers an accessible and standardized framework for integrating LLMs into behavioral research. By lowering technical barriers and enabling flexible, real-time interactions, it broadens the range of experimental designs that researchers can feasibly implement and helps ensure that studies involving LLMs can be compared and replicated with greater fidelity. As conversational AI becomes increasingly embedded in empirical research, tools like Simple Chat can contribute to transparency, reproducibility, and rigor. We anticipate that this framework will not only facilitate current investigations of human--LLM interaction, but also support the development of new methodological paradigms that harness the growing capabilities of modern language models.

\section{Declarations}

\subsection{Funding}
No funding was received to assist with the preparation of this manuscript.

\subsection{Conflicts of Interest}
The authors have no relevant financial or non-financial interests to disclose.

\subsection{Ethics Approval}
This manuscript introduces a tool and does not involve research with human participants.

\subsection{Consent to Participate}
This manuscript does not involve research with human participants.

\subsection{Consent for Publication}
This manuscript does not involve research with human participants.

\subsection{Availability of Data and Materials}
The latency benchmark scripts, raw data, and results reported here are available in the \code{benchmarks/latency} directory of the Simple Chat GitHub repository at commit \code{b1ad433}: \url{https://github.com/center-for-humans-and-machines/simple-chat/tree/b1ad433f50a3ff96ff39022cb60241fc3e051b2d/benchmarks/latency}. No data from human participants were collected.

\subsection{Code Availability}
The source code is available on GitHub at \url{https://github.com/center-for-humans-and-machines/simple-chat}.

\subsection{Authors' Contributions}
L.B. conceptualized the tool. R.B.S., A.D., L.B. provided software. R.B.S. wrote the draft of the manuscript. R.B.S., A.D., L.B. reviewed and edited the manuscript. R.B.S. provided visualization.

\section{Acknowledgements}

We thank Ali Alhosseini, Samira Fakhri, Omar Sherif, Jaeeun Shin, Bramantyo Supriyatno, and Tobias Werner for their help in developing and testing Simple Chat. We thank Iyad Rahwan for supporting the provision of Simple Chat as a Service. We are also grateful to Philip Jakob and Anastasia Kozyreva for their feedback on our manuscript.

\vspace{\baselineskip}
\section{Open Practices Statement}

Simple Chat is open-source and has already been published on GitHub at \url{https://github.com/center-for-humans-and-machines/simple-chat}, where its source code, documentation, and a tutorial video are publicly available. The Simple Chat versions of the materials used in the two case studies are also available on GitHub: the Qualtrics template in the Simple Chat repository and the oTree experiment at \url{https://github.com/center-for-humans-and-machines/simple-chat-otree-walter-2025}. This article does not report new studies with human participants; therefore, no participant data were collected and no studies were preregistered. The scripts and raw data of the latency benchmark are available in the Simple Chat repository. We welcome contributions and we aim to foster an active open-source community around Simple Chat.

\clearpage
\section{References}
\printbibliography[heading=none]

\clearpage
\appendix
\phantomsection\label{app:schema}
\addcontentsline{toc}{section}{Appendix A: Database Schema}
\centredline{\textbf{Appendix A}}
\centredline{\textbf{Database Schema}}
\FigureBlock{476.1pt}{12.2pt}{\hbox to \linewidth{\hss\resizebox{407pt}{458.7pt}{\singlespace\renewcommand{\sfdefault}{lmss}\renewcommand{\footnotesize}{\fontsize{8}{9.5}\selectfont}\input{figures/database-schema.tex}}\hss}}
\Note{The \codeV{search_parameters} field in the \code{SearchParameters} collection stores query parameters specified in \fix{Tab.~\ref{tab:params}}{Table~\ref{tab:params}}. $\ast$\codeV{SearchParameters} indicates that \code{MessageDocument} entries include \codeV{SearchParameters} fields. The arrows between collections indicate a relationship from an underlined entry to another collection.}

\clearpage
\phantomsection\label{app:stack}
\addcontentsline{toc}{section}{Appendix B: Technology Stack}
\centredline{\textbf{Appendix B}}
\centredline{\textbf{Technology Stack}}
\setcounter{table}{0}\renewcommand{\thetable}{B\arabic{table}}\renewcommand{\theHtable}{B\arabic{table}}
\FigureTitle{table}{tab:stack}{Technologies Used to Implement the Services of Simple Chat}
\FigureBlock{98pt}{6.9pt}{\hbox{\TableSetup
  \begin{tabular}{p{70pt}>{\raggedright\arraybackslash}p{376pt}}
    \toprule
    \HeadRow\centering Service & \centering\arraybackslash Technology \\
    \midrule
    \Row[12pt]Frontend       & TypeScript, bundled with Parcel; served by nginx in production. \\
    \Row[12pt]Backend        & Python with the FastAPI framework. \\
    \Row[12pt]Dashboard      & React, written in JavaScript; served by nginx in production. \\
    \Row[12pt]Database       & MongoDB, a NoSQL open-source database, run in a container by the Docker Compose file. \\
    \Row[12pt]Object storage & Any S3-compatible object storage, such as Amazon S3 or MinIO, configured in the \code{.env} file. \\
    \Row[12pt]Deployment     & Docker with the Docker Compose plugin. \\
    \bottomrule
  \end{tabular}}}
\Note{The services and their roles are described in Table~\ref{tab:services}.}

\end{document}

%% file: figures/database-schema.tex
  \tikzset{
    basic/.style={
      draw,
      rectangle split,
      rectangle split parts=2,
      rectangle split part fill={Salmon!20,white},
      minimum width=3cm,
      text width=3cm,
      align=left,
      font=\sffamily\footnotesize
    },
    arr/.style={-Stealth, thick},
    label/.style={font=\footnotesize\sffamily}
  }

  \begin{tikzpicture}[node distance=1.5cm and 2cm]

  \node[basic, minimum width=6.8cm, text width=6.7cm] (conversations) {Conversations
    \nodepart{second}
    \begin{tabular}{@{}ll}
    \_id: & str \\
    \underline{conversation\_id}: & str \\
    created\_at: & str \\
    \underline{participant\_id}: & str \\
    experiment\_id: & str \\
    model: & str \\
    messages: & list[\underline{MessageDocument}] \\
    \underline{project\_id}: & str \\
    \underline{custom\_system\_message\_id}: & str? \\
    \end{tabular}
  };

  \node[basic, left=of conversations] (participant) {Participant
    \nodepart{second}
    \begin{tabular}{@{}ll}
    \_id: & str \\
    \underline{participant\_id}: & str \\
    black\_listed: & bool \\
    \end{tabular}
  };

  \node[basic, below=of conversations] (project) {Project
    \nodepart{second}
    \begin{tabular}{@{}ll}
    \_id: & str \\
    \underline{project\_id}: & str \\
    created\_at: & str \\
    created\_by: & str \\
    system\_message: & str \\
    \end{tabular}
  };

  \node[basic, minimum width=4.7cm, text width=4.7cm, below=of project] (llmcall) {LLmCall
    \nodepart{second}
    \begin{tabular}{@{}ll}
    \_id: & str \\
    \underline{project\_id}: & str \\
    chat: & list[dict\{role, content\}] \\
    response: & dict \\
    created\_at: & str \\
    \end{tabular}
  };

  \node[basic, minimum width=6.9cm, text width=6.8cm, below=of participant] (msgdoc) {MessageDocument
    \nodepart{second}
    \begin{tabular}{@{}ll}
    \_id: & str \\
    content: & str \textbar{} list \\
    role: & "user" \textbar{} "assistant" \textbar{} "system" \\
    timestamp: & str \\
    type: & "text" \textbar{} "image" \\
    $\ast$SearchParameters: & SearchParameters
    \end{tabular}
  };

  \node[basic, above left=of conversations] (csm) {CustomSystemMessage
    \nodepart{second}
    \begin{tabular}{@{}ll}
    \_id: & str \\
    \underline{system\_message}: & str \\
    project\_id: & str \\
    created\_by: & str \\
    assistant\_first: & bool \\
    created\_at: & str \\
    \end{tabular}
  };

  \node[basic, minimum width=3.5cm, text width=3.4cm, above=of conversations] (searchparams) {SearchParameters
    \nodepart{second}
    \begin{tabular}{@{}ll}
    \_id: & str \\
    \underline{session\_id}: & str \\
    timestamp: & str \\
    search\_parameters: & dict \\
    \end{tabular}
  };

  \draw[arr] (conversations) -- (participant) node[label, midway, above] {participant\_id};
  \draw[arr] (conversations) -- (project) node[label, midway, right] {project\_id};
  \draw[arr] (conversations) -- (csm) node[label, midway, below left] {custom\_system\_message\_id};
  \draw[arr] (llmcall) -- (project) node[label, midway, right] {project\_id};
  \draw[arr] (llmcall) -- (msgdoc) node[label, midway, above right] {role};
  \draw[arr] (conversations) -- (msgdoc) node[label, midway, above left] {messages};

  \end{tikzpicture}